# Gravity driven deterministic lateral displacement for suspended particles in a 3D obstacle array

*Siqi Du and German Drazer\**

Mechanical and Aerospace Engineering Department

Rutgers, The State University of New Jersey

Piscataway, NJ.

## Abstract

We present a simple solution to enhance the separation ability of deterministic lateral displacement (DLD) systems by expanding the two-dimensional nature of these devices and driving the particles into size-dependent, fully three-dimensional trajectories. Specifically, we drive the particles through an array of long cylindrical posts, such that they not only move in the plane perpendicular to the posts as in traditional two-dimensional DLD systems (in-plane motion), but also along the axial direction of the solid posts (out-of-plane motion). We show that the (projected) in-plane motion of the particles is completely analogous to that observed in 2D-DLD systems. In fact, a theoretical model originally developed for force-driven, two-dimensional DLD systems accurately describes the experimental results. More importantly, we analyze the particles out-of-plane motion and observe that, for certain orientations of the driving force, significant differences in the out-of-plane displacement depending on particle size. Therefore, taking advantage of both the in-plane and out-of-plane motion of the particles, it is possible to achieve the simultaneous fractionation of a polydisperse suspension into multiple streams.

# Introduction

Deterministic lateral displacement (DLD) is a popular separation method in microfluidics that can effectively fractionate a polydisperse suspension of particles by driving it through a periodic array of obstacles[1]. In addition to size separation, DLD has been successfully applied in microfluidic systems to separate species with different shapes and deformability[2,3]. It was originally proposed as a flow driven, *passive* separation method but we have recently shown that *active*, force-driven DLD (f-DLD) is also effective in separating species by size[4–6] and shape[3]. In the majority of DLD systems the obstacle array is a periodic arrangement of cylindrical posts, but other obstacle shapes have been studied to enhance performance[7] or to separate non-spherical particles[8]. In all cases, however, the separation in DLD devices has been exclusively based on the motion of the suspended particles in the plane of the array, that is, the plane perpendicular to the cylindrical posts. As a result, and in spite of many variations[4,5,7], DLD systems have been limited to binary fractionations, in which a polydisperse sample is split into two streams, one of them usually containing the larger particles and the other containing the smaller ones.

Here, we propose a three-dimensional (3D) extension of DLD systems that overcomes the limitation of binary fractionation by taking advantage of the out-of-plane motion of the suspended particles. Specifically, we investigate an obstacle array with long cylindrical posts in which particles not only move *in-plane,* that is, perpendicular to the obstacles, but also *out-of-plane*, i.e. in the direction along the cylindrical obstacles. We have shown in previous work that macroscopic DLD models can facilitate detailed research on the particle motion inside the obstacle array[3,9,10]. Therefore, we designed a macroscopic setup that allows for direct visualization of the particles moving through the array of long cylindrical posts and to set at an arbitrary orientation with respect to the driving force (gravity). In this way, we are able to control the relative magnitude of the in-plane and out-of-plane components of the driving force. We perform experiments with particles of different sizes and for a wide range of force orientations with respect to the obstacle array. In all cases, we observe that the *in-plane motion* of the particles, that is the motion projected onto the plane of the array, is analogous to that found in two-dimensional (2D) DLD systems. In particular, there exists a transition from *locked mode* in which particles move in a principal direction of the array to *zigzag mode* in which they follow the external force more closely. Analogous to the 2D-DLD case, the fact that particles of different size transition from *locked mode* to *zigzag mode* at different orientations of the driving force is the basis for their in-plane separation. More importantly, we show that the out-of-plane motion of the particles is also size dependent. Therefore, 3D-DLD enables the simultaneous separation both in-plane and out-of-plane, thus increasing resolution and making it possible to fractionate a polydisperse suspension into multiple streams. In fact, based on our characterization experiments we demonstrate the simultaneous separation of particles of three different sizes coming out of our 3D-DLD system with excellent results.

## Materials and methods

*Experimental setup and materials*

A schematic view of the experimental setup is presented in Figure 1. The 3D array of obstacles is created using steel rods (diameter $D = 2$ mm, McMater-Carr Inc.) arranged in a square array between two parallel acrylic plates (see Figure 1a). The separation between rods in the array is $l = 6$ mm, and the separation between the acrylic plates is $L = 14$ cm. The two acrylic plates are fixed on a square acrylic base so that the obstacle array can be rotated as one solid object. The obstacle array is then placed on a supporting rectangular acrylic plate that can be tilted to an arbitrary angle $\theta$ with respect to a level surface (see Figure 1b). In addition, the base can be arbitrarily rotated an angle $\varphi$ with respect to the supporting plate, as shown in Figure 1c. The tilt angle, $\theta$, and the rotation angle, $\varphi$, allow us to control the orientation of the obstacle array with respect to gravity.

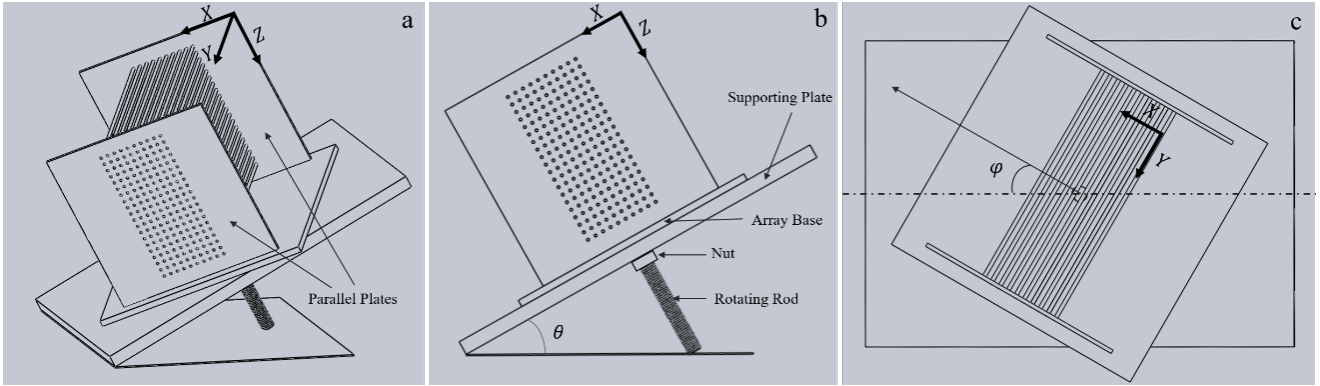

Figure 1 Schematic view of the experimental setup. a) Perspective view. b) Side view for a rotation angle $\varphi = 0°$. c) Top view of the rotating obstacle array on the supporting plate.

We then place our 3D-DLD system into a container filled with corn oil (viscosity $\mu = 52.3$ mPa · s, density $\rho_f = 0.926$ g/cm$^3$). We performed experiments covering tilt angles from 15.8° to 32.0°, and the rotation angle is varied (approximately) between 5° and 85°, depending on particle size. In a given experiment, with fixed slope and rotation angles, we release one particle at a time into the system, to eliminate particle-particle interactions. We use nylon particles with diameters $d = 1.59$ mm, $2.38$ mm and $3.16$ mm (McMater-Carr Inc.), and a total of 20-30 particles are tracked in each experiment. The density of the particles is $\rho_s = 1.135$ g/cm$^3$. The particle Reynolds number in our system is given by $\text{Re}_p = \frac{\rho_f U d}{\mu}$, where U is the characteristic velocity of the particles. The largest value, estimated using the average sedimenting velocity of the largest particles (U=3.6 mm/s), is $\text{Re}_p \sim 0.2$. The Stokes number is given by $\text{St} = \frac{1}{9}(\frac{\rho_s}{\rho_f})\text{Re}_p$, and the corresponding maximum value is thus estimated to be $\text{St} \sim 0.03$. We note that these values are consistent with those typically found in microfluidic systems.

*Problem geometry and coordinate system*

As shown in Figure 1a, the X and Z axes define the plane perpendicular to the obstacles, and the Y axis is taken as the direction parallel to the cylindrical posts (parallel to their axes). Figure 2a is a schematic representation of two typical trajectories followed by particles inside the 3D obstacle array, one corresponding to *zigzag mode* (small circles) and the other one corresponding to *locked mode* (large circles). Figure 2b shows the projection of the trajectories onto the XZ plane.

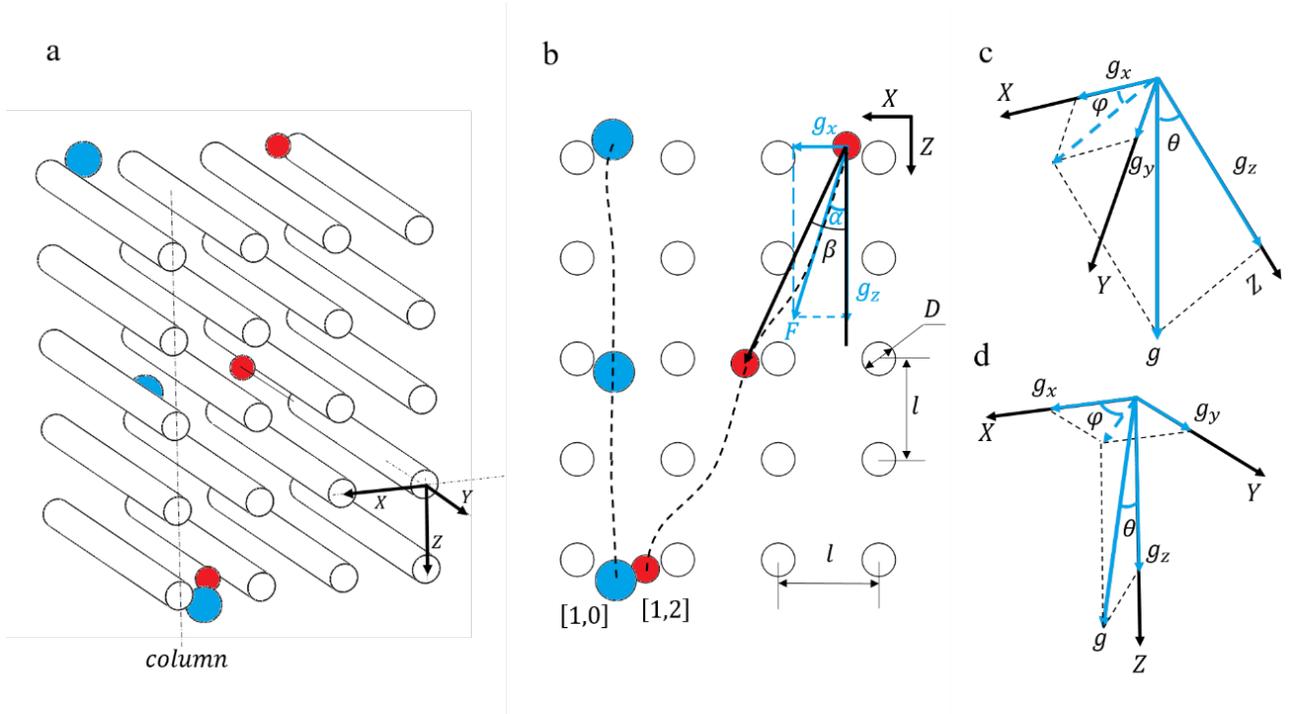

Figure 2: a) Typical particle trajectories. The smaller circles represent the trajectory of a particle moving inside the obstacle array in *zigzag mode* with a [1,2] periodicity, and the larger circles represent the trajectory of a particle moving in *locked mode*, i.e [1,0] periodicity. b) Projection of the trajectories shown in a) onto the XZ plane, indicating the forcing angle $\alpha$ and the migration angle $\beta$. c) Coordinate system of the setup viewed from the laboratory reference frame (gravity is pointing vertically downwards). d) Gravity force in the coordinate system of the setup.

When projected onto the XZ plane, particle trajectories can be compared to the 2D case. To this end, we determine the forcing angle in the XZ plane, $\alpha$, i.e. the angle between the in-plane projection of the force acting on the particles and the Z axis, and the migration angle in the XZ plane, $\beta$, i.e. the angle between the projected trajectory (onto the XZ plane) and the Z axis (see Figure 2b). The different components of the driving force (gravity) can be written in the terms of the slope angle $\theta$ and the rotation angle $\varphi$ as follows (see Figure 2b, 2c and 2d):

$$g_z = g\cos(\theta), \tag{1a}$$

$$g_x = g\sin(\theta)\cos(\varphi), \tag{1b}$$

$$g_y = g\sin(\theta)\sin(\varphi). \tag{1c}$$

The forcing angle in the XZ plane is given by

$$\tan(\alpha) = {g_x}/{g_z} = \tan(\theta)\cos(\varphi). \tag{2}$$

Note that for a fixed tilt angle $\theta$, the possible forcing angles that can be obtained by varying the rotation angle $\varphi$ are limited to $0 < \alpha < \theta$.

## Results and discussion

*Particle in-plane motion and comparison with 2D-DLD*

In previous work, we have shown that particles moving in *zigzag mode* have periodic trajectories. The periodicity of a trajectory is described by its average direction [p,q], where p, q are Miller indices. For example, in Figure 2b, the small circles represent a particle moving inside the obstacle array with periodicity [1,2]. Particles moving in *locked mode*, represented by the large circles in Figure 2b, move with periodicity [1,0]. In 2D-DLD, particles of all sizes were observed to transition from *locked mode* (periodicity [1,0]) to *zigzag mode* with a different periodicity, as the forcing angle increases from $\alpha = 0°$.[4,11] The angle at which the transition occur is defined as the critical angle $\alpha_c$ and, in principle, it is different for each type of particle.[11]

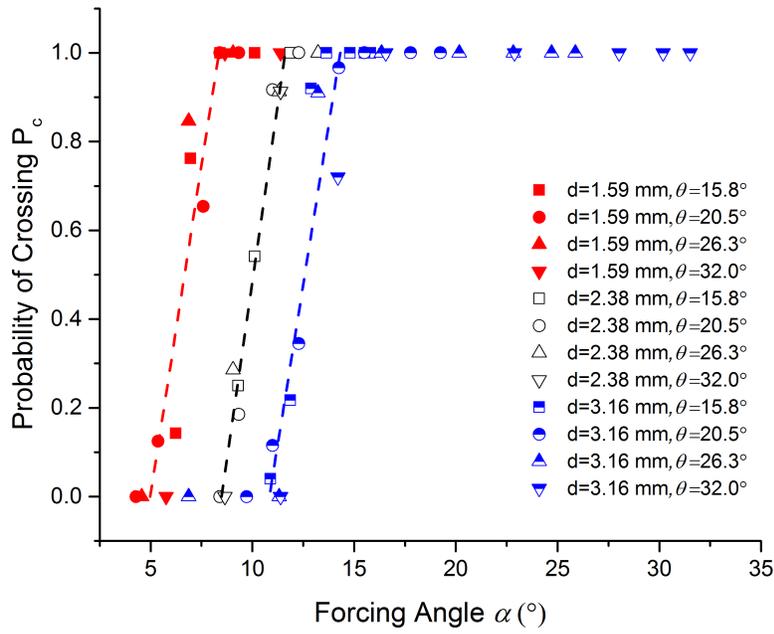

Figure 3 Probability of crossing as a function of the forcing angle. Solid, open and the half-solid symbols correspond to 1.59 mm, 2.38 mm and 3.16 mm particles, respectively. Different symbol shapes correspond to different slope angles as indicated.

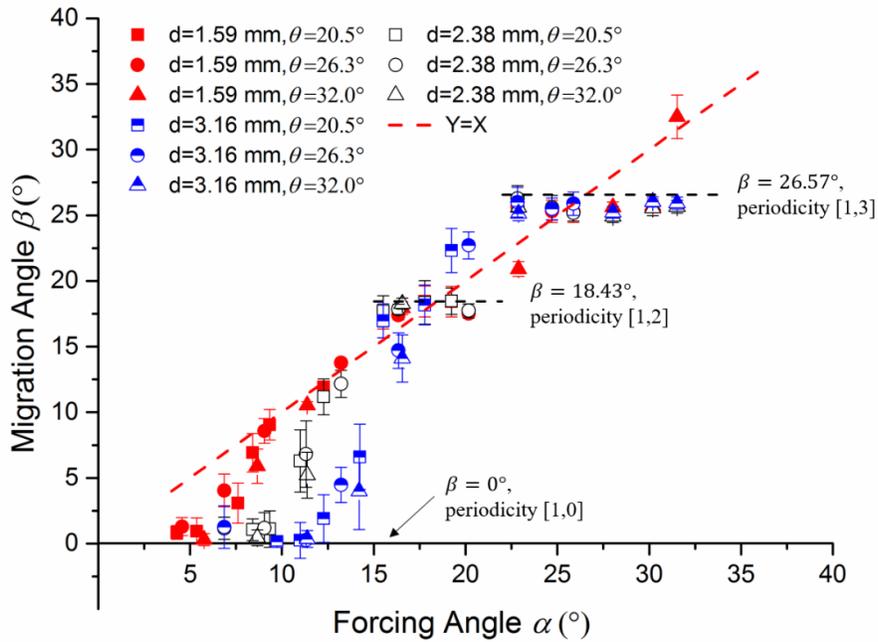

Figure 4 Migration angles as a function of forcing angle. Solid, open and the half-solid symbols correspond to 1.59 mm, 2.38 mm and 3.16 mm particles, respectively. Different symbol shapes correspond to different slope angles as indicated. The dashed line represents β = α. The directions [1, 0], [1, 2] and [1, 3] and the corresponding migration angles are indicated.

To investigate the *locked-to-zigzag* transition we introduce the probability of crossing, $P_C$, defined as the fraction of a given size of particles that move in *zigzag mode* out of the total number of those particles in a given experiment. In Figure 3, we plot $P_C$ as a function of the forcing angle for the different particles considered here. Consistent with 2D-DLD results, we observe sharp transitions in the crossing probability for all particle sizes. We estimate the critical angle $\alpha_c$ for each particle size as the forcing angle where its probability of crossing is equal to $1/2$, calculated using a linear fit of the intermediate $P_C$ values (see Figure 3). For 1.59 mm, 2.38 mm, and 3.16 mm particles the estimated values of the critical angle are 6.7° ± 1.7°, 10.0° ± 1.5° and 12.6° ±1.7°, respectively. Also analogous to the 2D case, the critical angle increases with particle size, which enables size-based separation. Additionally, we observe that for the same size of particles, the experimental results obtained with different tilt angles collapse into a single curve, which is consistent with the in-plane motion of the particles being independent of the motion along the posts. This is expected for the motion of a suspended particle past an array of posts at low Reynolds numbers, as long as particle-obstacle non-hydrodynamic interactions can be approximated by hard-core repulsion forces[12,13,11].

In Figure 4, we show the migration angle as a function of forcing angle for all the particles. As expected, for forcing angles smaller than the critical angle, the migration angle remains locked at $\beta = 0°$, i.e. particles are moving in *locked mode*. For forcing angles larger than the critical angle, particles migrate in *zigzag mode* with $\beta > 0°$. Again, we observe that the migration angle is independent of the tilt angle, which suggests that the in-plane motion of the particles is in fact independent from the out-of-plane dynamics. Figure 4, also shows that, when particles are moving in *zigzag mode*, their migration is not necessarily aligned with the forcing direction. In fact, Figure 4 shows clear 'plateaus' in the migration angle vs. forcing angle curves, indicating a constant migration angle for finite intervals of the forcing angle. This phenomenon, known as *directional locking*, is also present in the 2D case[14].

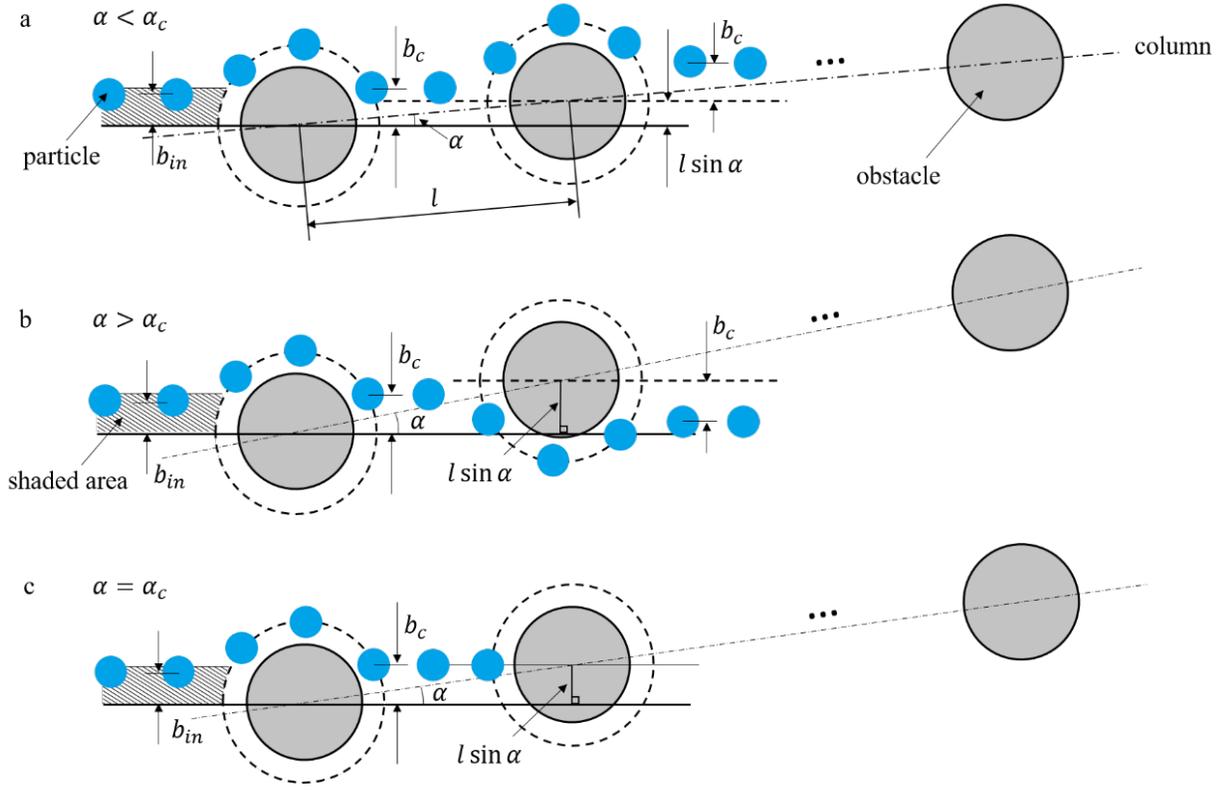

Figure 5: Schematic view of particle-obstacle irreversible collisions depending on the magnitude of the lateral shift between obstacles $l \sin \alpha$ compared to the critical impact parameter $b_c$. Note that collisions are irreversible, $b_{in} < b_c$ (shaded area), and particles come out of the interaction with the outgoing offset equal to the critical impact parameter $b_c$. a) A forcing angle such that $l \sin \alpha < b_c$, resulting in particles migrating in *locked mode*. b) A forcing angle corresponding to $l \sin \alpha > b_c$, which leads to particles migrating in *zigzag mode*. c) Forcing at the critical angle, i.e. $l \sin \alpha = b_c$.

*Migration model*

Let us then consider a model originally developed for 2D-DLD systems based on the assumption that a suspended particle only interacts with a single obstacle at a time (*dilute limit*). The trajectory of the particles is therefore determined by a sequence of individual particle-obstacle collisions[11,15,16]. For each individual particle-obstacle collision, the effect of all the short-range non-hydrodynamic repulsive forces between the particle and the obstacle is approximated by a hard-core potential. The hard-core repulsion prevents particles from coming closer to the obstacles than a given minimum separation but does not affect the particle trajectory otherwise. It is also important to note that, in the absence of inertia effects (at low Reynolds numbers) the minimum separation between the particle surface and the obstacle during a particle-obstacle collision is uniquely determined by the initial offset $b_{in}$ (see Figure 5). Therefore, for each particle size, we can define a critical initial offset $b_c$ as the initial offset leading to the minimum separation set by the hard-core repulsion. Then, collisions can be divided into two groups subject to the relation between $b_{in}$ and $b_c$. Collisions for which $b_{in} > b_c$, are *reversible*, particle trajectories are fore-and-aft symmetric and hence there is no lateral displacement after the suspended particle moves past the obstacle. On the other hand, collisions for which $b_{in} \leq b_c$ (shaded region in Figure 5) are *irreversible* and their outgoing offset is always $b_c$. That is, irreversible collisions result in a net lateral displacement of magnitude $|b_c - b_{in}|$. The fact that particles colliding with an obstacle with $b_{in} \leq b_c$, i.e. inside the shaded region in Figure 5, come out of the collision with the same offset $b_c$ results in directional locking.

Figure 5 shows three trajectories representative of the *locked-to-zigzag* transition according to the collision model just introduced. First, when the lateral displacement between two neighboring obstacles, $l \sin \alpha$, is less than $b_c$, as shown in Figure 5a, particles will be continuously displaced by successive obstacles due to irreversible collisions. That is, in this case particles will migrate in *locked mode*. In the figure, this corresponds to particles being displaced vertically up after each particle-obstacle collision and staying within a column of obstacles as indicated. On the other hand, when $l \sin \alpha > b_c$ (Figure 5b), particles coming out of an irreversible collision will cross through their original obstacle column, i.e they move in *zigzag mode*. The mode transition takes place when the driving force angle increases past its critical value, which depends on the particle-obstacle pair. A situation in which particles are driven exactly at the critical forcing angle is shown in Figure 5c. This corresponds to a particle coming out of an irreversible collision and heading into the next collision with $b_{in} = 0$, as shown in the figure, which explains the sharp nature of the transition. Given $b_c$, and assuming that successive collisions are independent, the model is able to predict the migration angle at any forcing angle. Therefore, and in addition to the set of critical angle values calculated from the crossing probability, we obtain a second estimate of the critical angles for the particles by fitting the average migration angles with the proposed model (where $b_c$ is the only fitting parameter). The results are plotted in Figure 6 and the two sets of $b_c$ values are presented in Table 1. We clearly observe the existence of directional locking for all sizes of particles and there is good agreement with the migration model.

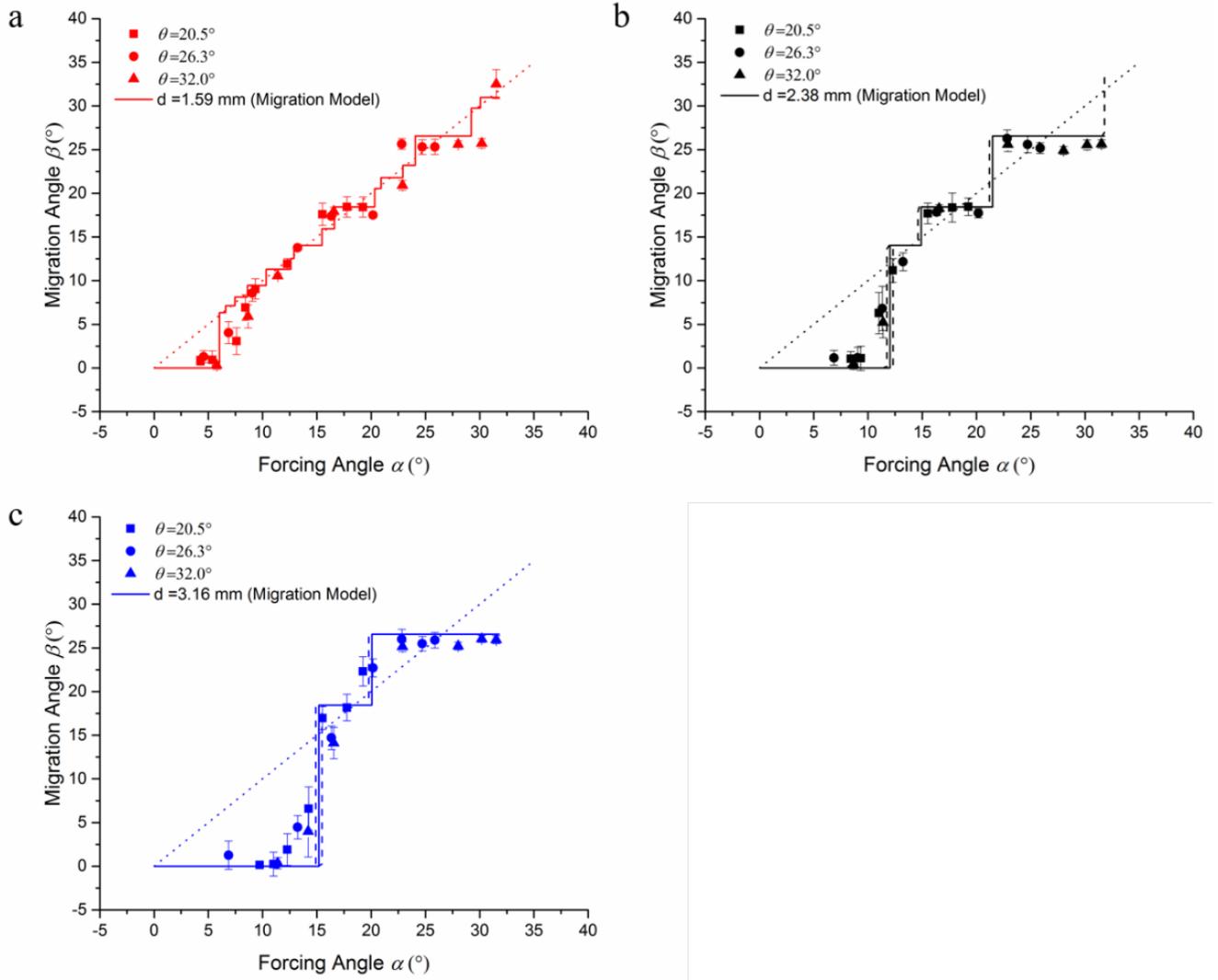

Figure 6: Migration angle as a function of forcing angle for (a) 1.59 mm particles, (b) 2.38 mm particles and (c) 3.16 mm particles. The solid line represents the best fit of the experimental results with the proposed model. The critical offset values obtained from the fit are $b_c = 0.61$ mm, $1.24 \pm 0.04$ mm and $1.57 \pm 0.03$ mm respectively. The dashed lines indicate the uncertainty of the fitting parameter $b_c$ in each plot. The dotted straight line indicates $\beta = \alpha$ for reference.

Table 1 Critical offset obtained from probability of crossing and model fitting

| Particle Size $d$ (mm) | $b_c$ from $P_c$ (mm) | $b_c$ from model (mm) |
|---|---|---|
| 1.59 | 0.70 ± 0.18 | 0.61 |
| 2.38 | 1.04 ± 0.15 | 1.24 ± 0.04 |
| 3.16 | 1.31 ± 0.17 | 1.57 ± 0.03 |

*Three-dimensional deterministic lateral displacement (3D-DLD)*

We consider the possible separative nature of the out-of-plane displacement $\Delta y$. In order to compare the displacement of different particles, as well as its dependence on the forcing direction, we normalize it by the in-plane displacement along the Z axis to obtain $\Delta y/\Delta z$. In , Figure 7 we present $\Delta y/\Delta z$ as a function of the in-plane forcing angle for all sizes of particles and for tilt angles $\theta$= 20.5°, 26.3° and 32.0°. As indicated in the plots, for all particle sizes, the out-of-plane displacement peaks around their individual critical angles. This suggests that the particle in-plane motion significantly affects the out-of-plane displacement. When the in-plane forcing angle is close to its critical angle, particles tend to stay close the obstacle longer, slowing down its in-plane-motion and resulting in a large out-of-plane displacement. As a result, we observe that for forcing angles $\alpha < 20°$, particles of different size can be separated by taking advantage of the differences in their out-of-plane displacement.

Finally, we demonstrate the simultaneous fractionation of all three sizes of particles by harnessing the out-of-plane separative displacement discussed above. To this end, we consider a forcing angle $\alpha \cong 12°$. According to Figure 3, with this forcing angle, the 3.16 mm particles migrate in *locked mode*, while the 2.38 mm particles and 1.59 mm particles migrate in *zigzag mode*. This results in the in-plane separation of the largest particles from the rest. On the other hand, the 2.38 mm and 1.59 mm particles migrate at the same angle and could not be separated based on the in-plane motion alone. This is, in fact, a typical situation in 2D-DLD systems, and usually limits the separation that can be performed to the binary fractionation of a complex suspension into two streams. On the other hand, Figure 7b, for example, shows that 1.59 mm and 2.38 mm particles would have a significant difference in their out-of-plane displacement, which enables their fractionation. In order to demonstrate the advantages of 3D-DLD we have also quantified the quality of this test separation. To this end, we added a collector at the bottom of our experimental setup. The collector is partitioned into three sections, which based on our previous experimental data, would collect each size of particles separately. The results are provided in terms of $n_{\alpha\beta}$ the number of particles of type α in the collection bin designed to capture particles of type β. We can then define the *efficiency* of the separation of particles of a given type as the fraction of such particles in the corresponding collection bin, $e_\alpha = n_{\alpha\alpha}/\sum_\beta n_{\alpha\beta}$, and the *purity* of the separation of particles of a given type as the fraction of particles of this type in the corresponding bin, $p_\alpha = n_{\alpha\alpha}/\sum_\beta n_{\beta\alpha}$.

We first perform experiments by releasing one particle at a time into the device, in order to avoid particle-particle interactions and the results are presented in Table 2. We obtain excellent separation results, with efficiencies ≥ 95% and purities ≥ 89%. Then, in order to increase the throughput of the separation, we performed exploratory experiments introducing a mixture of 3-6 particles of different sizes at the same time and the results are presented in Table 3. Although both efficiency and purity values are still reasonably good, a clear reduction is observed, which suggests that further experiments are needed to investigate throughput limitations of the proposed system. On the other hand, we have recently demonstrated the use of centrifugal force to drive a 2D-DLD system[6], an alternative that could also benefit from the proposed 3D approach and would inherently lead to higher throughputs.

Table 2 Separation results in the absence of particle-particle interaction

| Particle Size / Bin Number | 1.59 mm | 2.38 mm | 3.16 mm | Purity |
|---|---|---|---|---|
| 1 | 1 | 1 | 17 | 89% |
| 2 | 0 | 21 | 0 | 100% |
| 3 | 24 | 0 | 0 | 100% |
| Efficiency | 96% | 95% | 100% | |

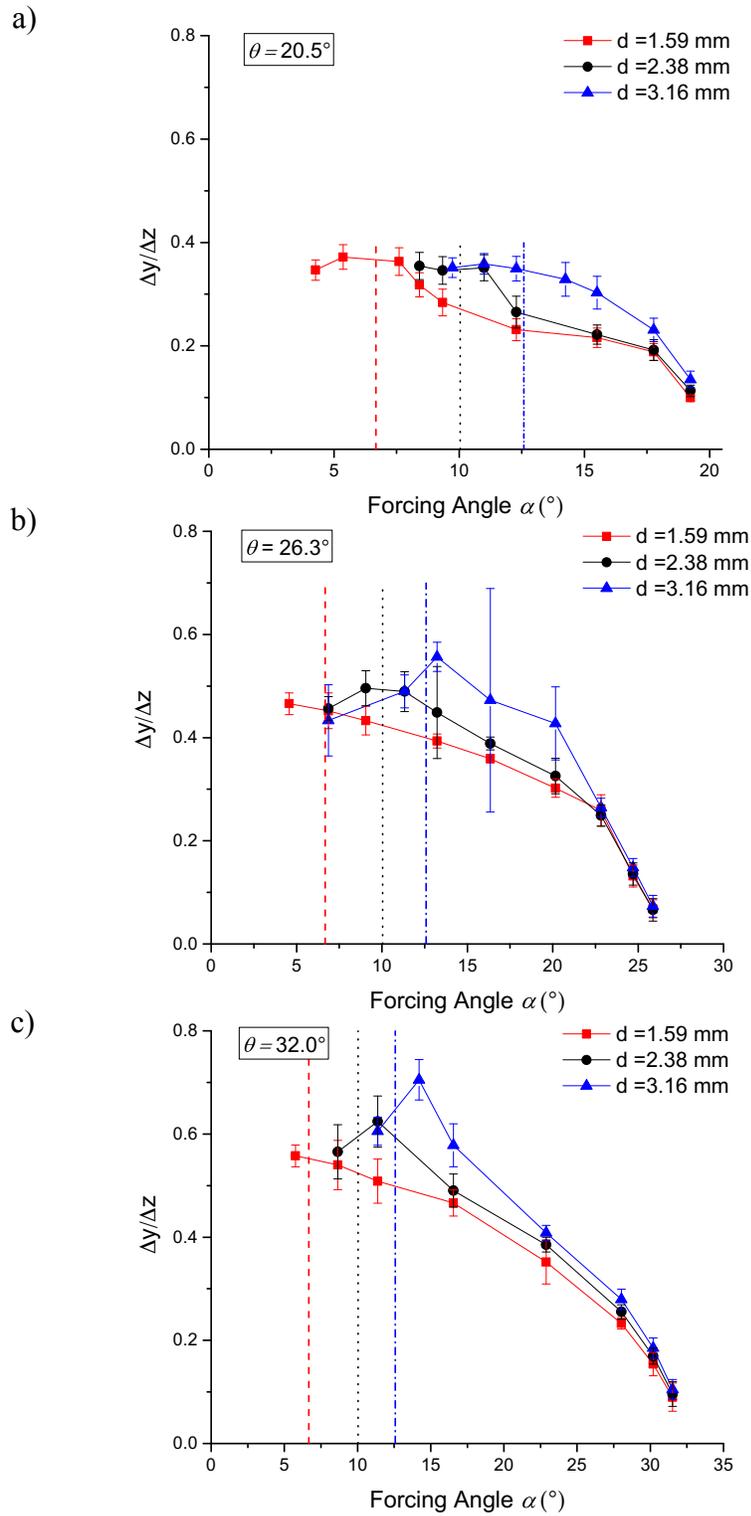

Figure 7: Out-of-plane displacement as a funtion of the in-plane forcing angle for tilt angles: (a) 20.5°, (b) 26.3° and (c) 32.0°. The dashed, dotted and dot dashed vertical lines in each plot represent the critical angles (obtained from the probability of crossing) for 1.59 mm particle, 2.38 mm particle and 3.16 mm particle, respectively.

Table 3 Separation results in the presence particle-particle interaction

| Particle size / Bin number | 1.59 mm | 2.38 mm | 3.16 mm | Purity |
|---|---|---|---|---|
| 1 | 0 | 1 | 17 | 94% |
| 2 | 5 | 22 | 0 | 82% |
| 3 | 18 | 2 | 0 | 90% |
| Efficiency | 78% | 88% | 100% | |

# Conclusions

We present a simple concept to enhance separation in DLD systems, based on extending the traditionally 2D method into the third dimension by using an array of long cylindrical posts. First, we demonstrated that when projected onto the plane of the array, the particle in-plane migration pattern is analogous to the force-driven 2D-DLD case. We observed the existence of a *locked mode* when the forcing angle is relatively small, and a sharp transition into *zigzag mode* when the forcing angle is increased past a critical value (critical angle). The fact that the critical angle depends on particle size enables the in-plane fractionation. We also observed that the particles in-plane trajectories are independent of the tilt angle and, therefore, independent of the out-of-plane motion. More important for separation, we observed that the particle out-of-plane displacement does depend on the in-plane motion, with the largest displacements for each type of particle observed when the forcing angle is close to the corresponding critical value. Therefore, the differences in critical angle with particle size not only enable in-plane separation but also lead to different out-of-plane displacements that can be harnessed to enhance the separation ability of DLD systems. Based on such observation, we then demonstrated that a polydisperse suspension containing three different sizes of particles can be fractionated into its individual components using the proposed 3D-DLD system, with excellent efficiency and purity. Finally, we note that increasing separation throughput lead to a reduction in separation quality and further experiments are needed to explore the effect of particle-particle interactions in the proposed 3D-DLD system.

## Acknowledgement

This research is partially funded by the National Science Foundation Grant No. CBET-1339087.

**Contributions**

S.D. and G.D. conceived the experiments. S.D. built the experimental setup and conducted the experiments. S.D. and G.D. analyzed the results. Both authors wrote and reviewed the manuscript.

**Competing interests**

The authors declare no competing financial interests.

181818